# Quantum superconducting diode effect with perfect efficiency above liquid-nitrogen temperature

Heng Wang[1,2,3,#], Yuying Zhu[1,4,#,*], Zhonghua Bai[2], Zhaozheng Lyu[4,5], Jiangang Yang[5], Lin Zhao[5], X. J. Zhou[5], Genda Gu[6], Qi-Kun Xue[1,2,3,7,*], and Ding Zhang[1,2,4,7,8,*]

[1]Beijing Academy of Quantum Information Sciences, Beijing 100193, China

[2]State Key Laboratory of Low Dimensional Quantum Physics and Department of Physics, Tsinghua University, Beijing 100084, China

[3]Southern University of Science and Technology, Shenzhen 518055, China

[4]Hefei National Laboratory, Hefei 230088, China

[5]Beijing National Laboratory for Condensed Matter Physics, Institute of Physics, Chinese Academy of Sciences, Beijing 100190, China

[6]Condensed Matter Physics and Materials Science Department, Brookhaven National Laboratory, Upton, New York 11973, USA

[7]Frontier Science Center for Quantum Information, Beijing 100084, China

[8]RIKEN Center for Emergent Matter Science (CEMS), Wako, Saitama 351-0198, Japan

\# These authors contributed equally.

*Email: zhuyy@baqis.ac.cn

qkxue@mail.tsinghua.edu.cn

dingzhang@mail.tsinghua.edu.cn

## Abstract

The superconducting diode is an emergent device that juggles between the Cooper-paired state and the resistive state with unpaired quasiparticles. Here, we report a quantum version of the superconducting diode, which operates solely between Cooper-paired states. This type of quantum superconducting diode takes advantage of quantized Shapiro steps for digitized outputs. The devices consist of twisted high-temperature cuprate superconductors, and exhibit the following distinguished characteristics: (1) a non-reciprocal diode behavior can be simply initiated by current training without applying an external magnetic field; (2) perfect diode efficiency is achieved under microwave irradiations at a record-high working temperature; (3) the quantized nature of the output offers high resilience against input noises. These features open up unprecedented opportunities toward developing practical dissipationless quantum circuits.

The superconducting diode (SD) effect [1-5] manifests itself as asymmetric current-voltage characteristics, which is analogous to the semiconductor diode (Fig. 1a) but with an exchanged voltage-current relation (Fig. 1b). The critical supercurrent flowing in one direction, $I_c^+$ for example, is prominently smaller than that in the opposite direction ($I_c^-$). Consequently, the polarity of a current $I$ with its amplitude between $I_c^+$ and $|I_c^-|$, i.e., $|I_c^-| < |I| < I_c^+$, can switch the system between the conventional resistive state with quasiparticles and the dissipationless state of Cooper pairs. The SD effect helps unveil exotic physics at play, such as finite-momentum Cooper pairing [6], spontaneous time-reversal symmetry breaking [7], etc. Moreover, it constitutes a superconducting switch, showing potential for the next-generation cryogenic electronics [8-10]. In this regard, an ideal SD should maximize the asymmetry such that the critical current exists only in one direction (it approaches zero in the opposite direction) [11-14]. The diode efficiency $\eta$, which is defined as $\eta = \left||I_c^+| - |I_c^-|\right|/(I_c^+ + |I_c^-|)$, would then become 100%. The latest development [14,15] by utilizing the AC Josephson effect pointed to a facile route, among other proposals [12,13,16,17], toward realizing the ideal SD. However, the working temperature of the reported device [15] remained in the sub-Kelvin regime, and more critically, required a fine-tuned magnetic field. Lifting these lingering constraints is crucial but challenging toward practical SD.

In this work, we report field-free SDs that function even above liquid nitrogen temperature. The diode efficiency $\eta$ can reach 100% upon microwave irradiation. Most importantly, our devices, Josephson junctions made of twisted cuprate superconductors, take advantage of asymmetrically developed Shapiro steps, as schematically shown in Fig. 1c. The output voltage of our SDs can be conveniently switched by current polarity between zero and the quantized values of $nhf/2e$, where $f$ is the microwave frequency and $n$ is an integer, corresponding to two distinct states that are both formed by Cooper pairs. We demonstrate that this quantum superconducting diode (QSD) effect is superior to the usual SD effect for noise-resilient signal processing.

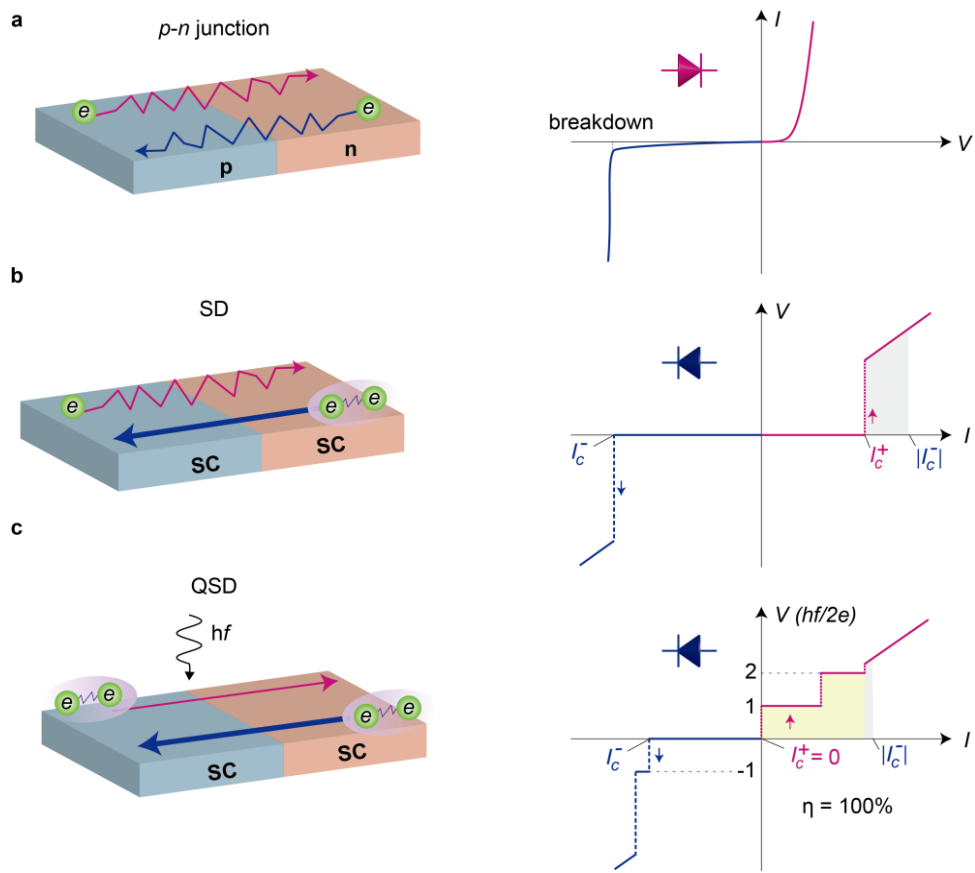

*Fig. 1 Semiconductor diode, Superconducting diode (SD) and Quantum Superconducting Diode (QSD).* **a,** Schematic drawing of a p-n semiconductor junction (left) and its typical $I$-$V$ characteristic (right). Current transport is carried by quasi-particles with dissipation (indicated by zigzag paths on the left). **b,** Schematic drawing of a SD (left) and its $I$-$V$ characteristic (right). Here, the critical current in the negative direction ($I_c^-$) is larger than that in the positive direction ($I_c^+$). In the asymmetric regime (gray region in the right panel), the supercurrent of Cooper pairs flows in one direction (blue arrow in the left panel) whereas unpaired quasi-particles flow in the opposite direction (purple arrow). **c,** Schematic drawing of a QSD under microwave irradiation and its $I$-$V$ characteristic with perfect efficiency. Plateaus in the $I$-$V$ curve reflect the Shapiro steps. Only the first two Shapiro steps are shown here for illustration purpose. In the regime highlighted by yellow, Cooper pairs flow in both directions (illustrated on the left) but a quantized voltage occurs in the positive direction.

The inset of Fig. 2a illustrates our device—a *c*-axis twisted Josephson junction made of two cuprate flakes exfoliated from either Bi$_2$Sr$_2$CaCu$_2$O$_{8+\delta}$ (Bi2212) or Bi$_{2-x}$Pb$_x$Sr$_2$CaCu$_2$O$_{8+\delta}$ (Pb-Bi2212) single crystals (Fabrication details given in Methods and in our previous works [18-20]). The twist angle of device A is 45°. We carry out all measurements in the absence of magnetic fields ($B=0$ T). Our twisted junction exhibits high transition temperature $T_c$ of about 81 K. Before the junction is initiated to show the SD effect, it hosts a critical Josephson current density $J_c$ of 286 A/cm² at $T=70$ K.

The *I-V* characteristic of device A, immediately after cooling from $T>T_c$, shows a nearly symmetric behavior (Fig. 2a). To introduce the SD effect, we send current pulses to the sample at the same temperature (70 K). A typical waveform of the current pulse is illustrated in the inset of Fig. 2b. After the current training, the *I-V* characteristic shows prominent asymmetry (Fig. 2b): the branch in the positive direction (purple) hosts a much lower critical current ($I_c^+$) than the one in the negative direction (blue). This asymmetry ($I_c^+ < |I_c^-|$) persists in the subsequent sweeps, such that the histograms of $I_c^+$ and $|I_c^-|$ (right panel of Fig. 2b) show distributions around sharply different current values. It reflects the on-demand formation of a SD at a rather high temperature of 70 K. The corresponding diode efficiency $\eta$ is about 30%, prominently larger than that achieved previously at the same temperature [21]. We defer further discussions on the mechanism of this SD effect to the end of this report.

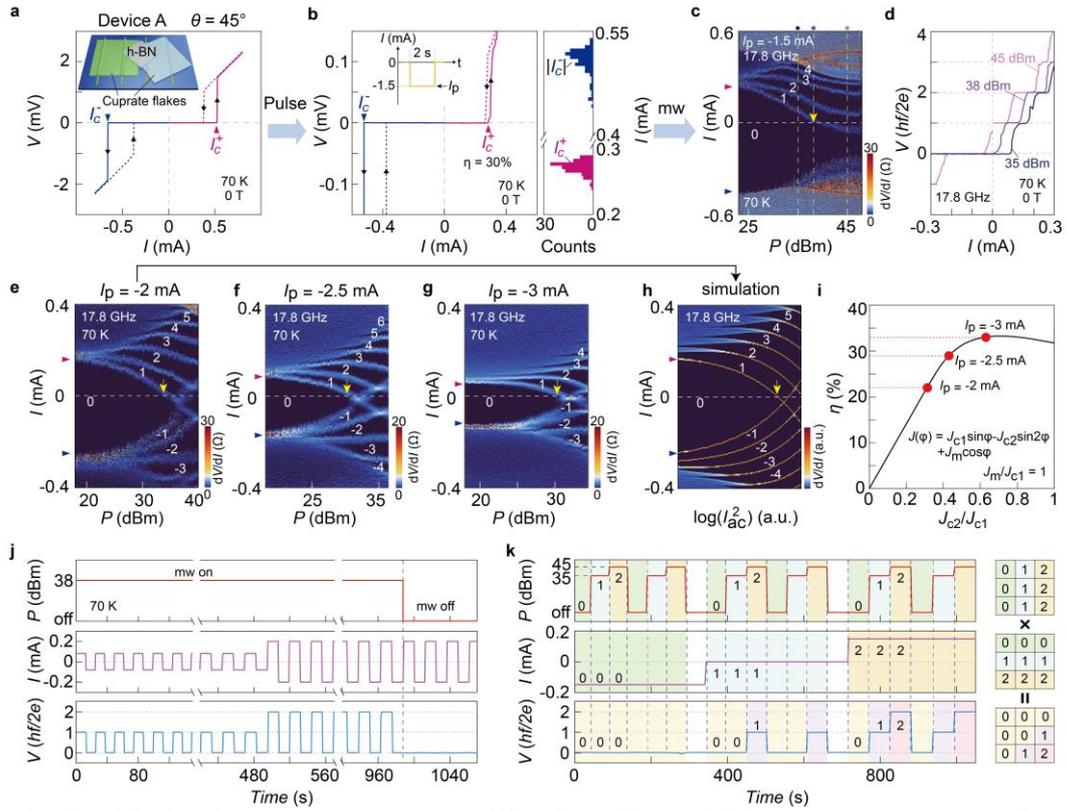

*Fig. 2 Initialization protocol and rectification effect of QSD.* All data are taken from device A at 70 K, 0 T. **a,** $I$-$V$ characteristic in the initial state. Arrows on the curves indicate the sweeping directions. Inset: Schematic drawing of the twisted junction. **b**, $I$-$V$ characteristic after sending a current pulse to the sample at 70 K. Inset: typical waveform of the current pulse. Right: histogram of $I_c^+$ and $|I_c^-|$. **c,** Differential resistance $dV/dI$ as a function of dc current $I$ and microwave power $P$ (microwave frequency: 17.8 GHz). Purple and blue arrows on the left abscissa indicate $I_c^+$ and $|I_c^-|$ at the lowest $P$. Yellow arrow in the panel indicates the point where $\eta$ reaches 100%. Integer numbers indicate the sequence of Shapiro steps. **d**, $I$-$V$ characteristics at selected microwave powers (indicated by dashed lines in panel c) at 17.8 GHz. **e-g,** Color-coded plots of $dV/dI$ as a function of $I$ and $P$ after applying different current pulses. Arrows on the left abscissa indicate $I_c^+$ and $|I_c^-|$ at the lowest $P$. Integer numbers indicate the sequence of Shapiro steps. **h**, Theoretical modelling of the asymmetrically developed Shapiro steps. Parameters are adjusted to simulate the experimental pattern in panel e. **i,** theoretically calculated $\eta$ as a function of $J_{c2}/J_{c1}$ (curve). Filled circles mark values of $\eta$ obtained experimentally after applying different current pulses (microwave off). **j,** Half-wave rectification controlled by microwave. Top: microwave on/off as a function of time. Middle/Bottom: input current/output voltage as a function of time. **k,** Demonstration of a modified AND operation by using the QSD effect. Top/middle/bottom: microwave power/input current/output voltage as a function of time.

Once the SD effect is initiated, we apply the microwave irradiation to drastically enhance the asymmetry. Figure 2c presents the evolution of current-voltage characteristics ($dV/\mathrm{d}I$-$I$) with increasing microwave power $P$. The dark blue area around $I=0$ represents the dissipationless region ($dV/\mathrm{d}I = 0$), i.e., current flows without generating any voltage ($V=0$). It is bounded by blue/dark red stripes in the positive/negative current direction (pointed by purple and blue arrows), reflecting the corresponding critical currents at a certain $P$. With increasing $P$, the region with $V=0$ (dark blue) shrinks but the positive section (with $I_c^+$) gets reduced more quickly[14]. Such an effective tuning progressively promotes the diode efficiency $\eta$, eventually leading to $\eta$=100% when a threshold power is exceeded (38 dBm, marked by the yellow arrow in Fig. 2c). We also note that a finite voltage at zero current bias can be registered at large $P$ (Fig. 2d). Such a wireless rectification effect is another manifestation of the diode behaviour [22,23]—similar to the photo-voltaic effect of a semiconductor diode. Apart from perfect efficiency, the color-coded plot also shows prominent Shapiro steps, corresponding to pajama stripes in Fig. 2c and plateaus at $nhf/2\mathrm{e}$ ($n=$ 1-3) in Fig. 2d.

In Fig. 2e-g, we demonstrate the tunability of the asymmetry by varying the amplitude of the current pulse $I_p$. At a fixed temperature of 70 K, we apply consecutively larger current pulse and measure the resulting $I$-$V$ characteristics. At $|I_p| > 1.5$ mA, the $I$-$V$ characteristics in both current directions show strongly suppressed hysteresis and the sample under microwave irradiation shows several Shapiro steps in both directions. The diode efficiency (microwave off) increases consecutively from about 22% at $I_p = -2$ mA to 34% at $I_p = -3$ mA. The diode efficiency (microwave on) reaches 100% at a lower microwave power with increasing $I_p$ (yellow arrows in Fig. 2e-g). The asymmetry imposed by the current pulse is reversible upon thermal cycling, excluding current induced damages as the origin for the SD effect. By warming up to a temperature above $T_c$ and cooling down again, we can bring the sample back to the initial state with the symmetric $I$-$V$ characteristic.

After recovering the symmetric behavior, the SD effect can be reintroduced by repeating the current pulsing procedure.

The SD effect together with the asymmetrically developed Shapiro steps (Fig. 2e for instance) can be modelled by considering a damped Josephson junction [14]. Figure 2h shows that the modelling yields the evolution of the Shapiro steps that well matches the experimental observation. In these simulations, we employ a current-phase relation (CPR) that takes into account an emergent second harmonic term and the effect of time reversal symmetry breaking (TRSB) (Methods). As shown in Fig. 2i, the diode efficiency $\eta$ can be tuned by increasing the ratio of $J_{c2}/J_{c1}$, where $J_{c1}$ and $J_{c2}$ represent the first and second harmonic terms in the CPR. This modelling indicates that applying a larger current pulse can effectively enhance $J_{c2}/J_{c1}$.

As shown in the color plots of Fig. 2c, e-g, the Shapiro steps develop asymmetrically between the positive and negative current directions. Such asymmetry is the defining character of a quantum superconducting diode (QSD) (Fig. 1c). Figure 2j showcases the half-wave rectification by choosing the amplitude of the input current (middle row) to be in the QSD regime. With the microwave on, the output signal (bottom row) alternates either between two states with $V = 0$ and $V = hf/2$e, or between the states with $V = 0$ and $V = hf/$e. The quantized output can be reliably controlled by the amplitude of the input signal. In contrast to the rectification of SD, here the states in both current directions are carried by Cooper pairs. Moreover, the half-wave output is immediately terminated once the microwave is turned off. The output resumes once the microwave irradiation is on (Fig. 2k). With a fixed input square wave, the output signal can be also controlled by tuning the microwave to different frequencies. Such a rapid microwave control is another advantage of QSD over SD.

In Fig. 2k, we realize a modified version of AND operation based on our QSD. We use the microwave and current as the two input channels. Each input channel hosts three states denoted as 0, 1, and 2 (indicated by numbers in Fig. 2k top and middle rows). By sending these signals into our QSD, the output signal yields 0, $hf/2e$, and $hf/e$, corresponding to the logic outputs of 0, 1, and 2. The right panels of Fig. 2k summarize the calculation table.

The QSD is much stable against noise from the input. In comparison, the output signal of a usual SD would fluctuate in the same manner as the input amplitude. In fact, the output signal of an SD jiggles even in the absence of any input noise, due to stochastic processes of the Josephson tunneling. This intrinsic fluctuation is shown in Fig. 3a. The left panel illustrates the input square wave while the middle panel shows the rectified output. The top part of the output wave shows clear uneven features. The fluctuation has a full-width at half-maximum (FWHM) that is 12.7% of the average value, as shown in the right panel of Fig. 3a. This FWHM corresponds to 7.47 μV, which is much larger than the instrumental noise $\sigma_{instru}$ of about 0.23 μV (Methods).

For the QSD, the output is unaffected by the stochastic process of Josephson tunneling. As demonstrated in Fig. 3b, the output signal (middle panel) stays flat and the histogram shows a much narrower distribution. Quantitatively, the FWHM is 0.23 μV (0.3% of the quantized output). It indicates that the output fluctuation is now limited only by the instrumental noise $\sigma_{instru}$. Figure 3c presents the data with the input fluctuations of $\pm 10\%$ of the set current amplitude $I_s$. Interestingly, the output voltage remains flat (middle panels) and the histogram of the non-zero output voltage remains narrow, irrespective of the input noise. We measured that the tolerance of input noise can be as high as $\pm 20\%$ of the input signal, where the current amplitude is in the middle of the Shapiro step.

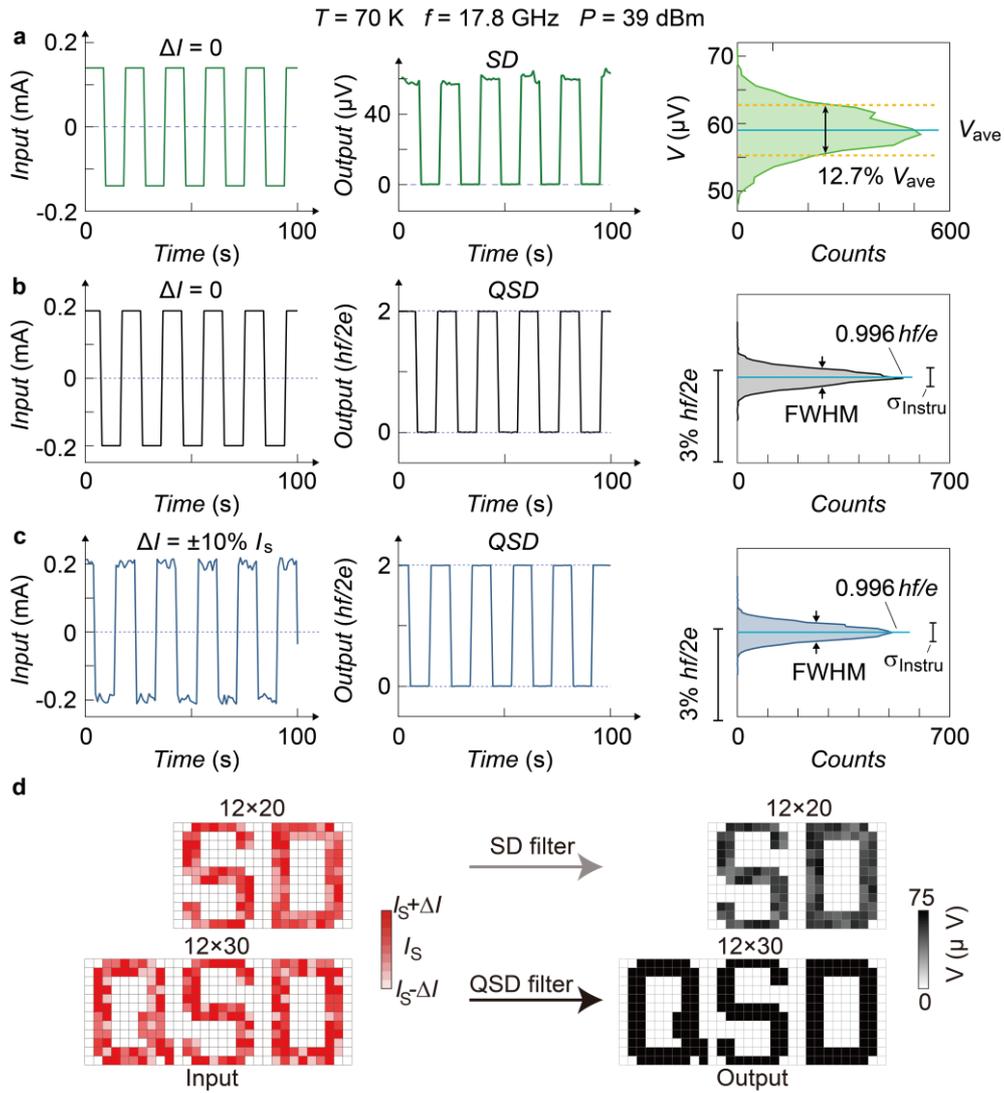

**Fig. 3 Noise suppression of QSD.** All data are collected from device A at 70 K. **a,** Representative output (middle) with the square wave input (left) in the SD regime. Right panel: histogram of the output over 500 repetitions. Vertical arrows indicate the full width at half maximum (FWHM). Solid horizontal line indicates the averaged value of the output signal in half of the period. **b, c,** Representative input waveforms (left), correspondent outputs (middle), and histograms of the output in the QSD regime without (**b**) or with (**c**) noise introduced to the input: $\pm 10\% I_s$, where $I_s$ stays in the QSD regime. Each histogram is taken for data over 500 repetitions. **d,** Applications of the SD and QSD as filters in signal processing. Left: letters of "S", "D" (top) and "Q", "S", "D" (bottom) as input signals. Each letter is depicted by 12 × 10 pixels. Empty pixels correspond to negative input current values. Pixels with the color scale correspond to input current in the range of $[I_s - \Delta I, I_s + \Delta I]$, where $\Delta I = 10\% I_s$. Here, $I_s = 140\ \mu A$ for the top SD filter and $I_s = 200\ \mu A$ for the bottom QSD filter. The current fluctuation was randomly chosen for each filled pixel.

Figure 3d visualizes the noise filtering effect by contrasting the output signals in the usual SD regime (top) and those in the QSD regime (bottom). Here, each letter is depicted in a grid of 12 by 10 pixels. For the input signal, the fully white pixels correspond to negative current values, while the pixels with color scales correspond to current in the range of $[I_s - \Delta I, I_s + \Delta I]$. Clearly, the output signal in the QSD regime (lower right of Fig. 3d) has much better contrast, indicating substantially enhanced signal to noise ratio of the device.

Our protocol can reliably induce the QSD effect in the twisted cuprate junctions. In Fig. 4a-e, we show another five QSDs with $\eta = 100\%$. The exact critical current varies because of the different sample size. Notably, data in Figure 4a, 4c and 4f are taken at or above liquid-nitrogen temperature, attesting to the unprecedented high working temperature of our devices. Also, device F exhibits clear Shapiro steps at 1/2 and 3/2, demonstrating the presence of higher-harmonic term in the current-phase relation (CPR) of the junction. Similarly, we observe half-integer steps in devices B and C with $\theta = 25°$ and $30°$.

Figure 4g summarizes the diode efficiency as a function of $\theta$ among 18 devices we measured in this study. Each empty symbol represents an individual device. By applying the current pulse, we can realize the SD effect with $\eta$ in the range of 17% to 40% (microwave off). In particular, $\eta$ (microwave off) reaches 37.5% at a temperature above the boiling point of liquid nitrogen—83 K—for device D, showing potential for practical energy-saving devices. The symbols connected by vertical lines correspond to samples tuned by microwave. Perfect diode efficiency can always be achieved by microwave irradiations as long as the SD effect is initiated. Figure 4g also includes several data points at $15°$, $25°$ and $30°$ with $\eta \sim 0\%$: two samples at each angle. These samples host symmetric $I$-$V$ characteristics that are unaffected by applying the current pulses. By contrast, the samples with $\theta = 40°$ and $45°$ (in total seven samples) can be easily tuned to become SDs. The yield of current pulse induced SD effect seems to become higher as the twist angle approaches $45°$. The implication of this angular dependence will be addressed in the discussion section.

In Fig. 4h, we provide an overview of the SDs made of various materials. Most of the reported SDs [1,4,6,7,21,24-43] only work in the low temperature regime with $\eta$ less than 50%. In addition, a well-adjusted magnetic field is often required for inducing the asymmetric $I$-$V$ characteristic. Using high temperature cuprate superconductors is a viable route toward a more practical SD. Our work unveils that this route has the benefit that well-developed Shapiro steps can be achieved even above liquid nitrogen temperature, facilitating the realization of QSD. Further improvement can be gained by employing cuprate superconductors with an even higher $T_c$, for instance the compound with copper oxide trilayers [44].

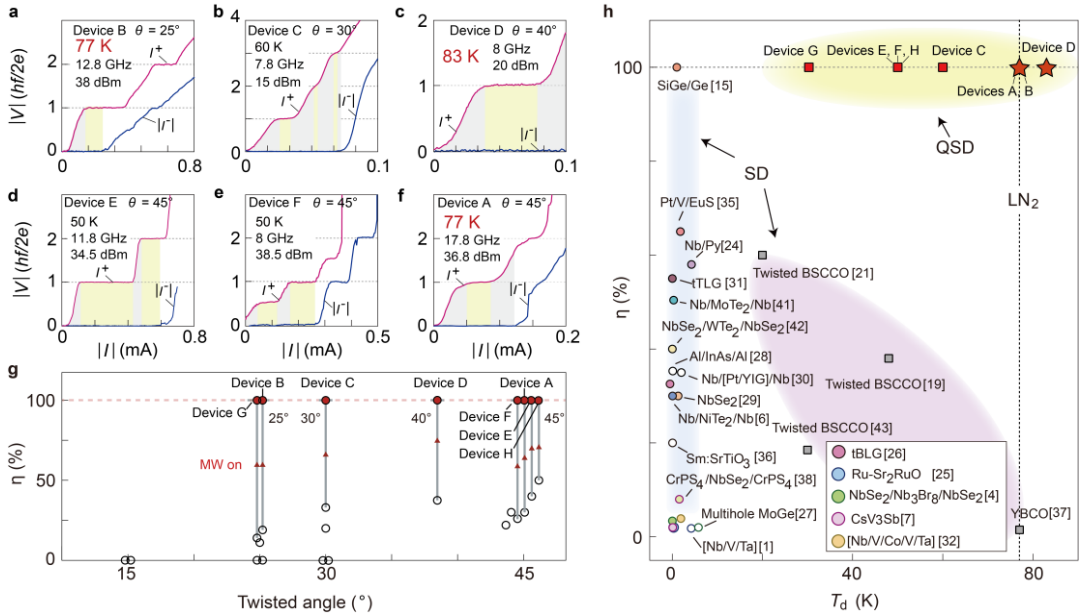

*Fig. 4 Overview of QSDs.* **a,b,c,d,e,f,** QSD effects realized in six different devices. High working temperatures for devices A, B, and D are marked by bold characters. **g,** Diode efficiency as a function of twist angle between the cuprate flakes. Each empty symbol corresponds to one device that we fabricated and measured. Vertical bars connecting to filled symbols represent realization of perfect efficiency by applying microwave. Data points for devices at 45 degrees are slightly offset horizontally for clarity. Data points at 15, 25, and 30 degrees are also horizontally offset. **h,** Diode efficiency and working temperature $T_d$ of the previously reported SDs[1,4,6,15,19,21,24-32,35-38,41-43] and our QSDs. Vertical line marks the boiling point of liquid nitrogen (LN$_2$).

The SD effect observed in twisted cuprates has been attributed to either the spontaneous TRSB [45,46] or the explicit TRSB [46]. The latter scenario may stem from magnetism [46] or vortices [19,21]. We point out several distinct differences between our experimental results and the theoretical proposal based on spontaneous TRSB. First, we realize the SD effect in multiple junctions at 45° [the error bar in θ is within 0.1° (Methods)] with $\eta$ reaching as large as 40% and at temperatures close to $T_c$. By contrast, the theory predicts that η drops below 30% as θ exceeds 44° and reaches zero at θ = 45°. In addition, the theoretical scenario is valid only for $T < 0.5T_c$; Secondly, we obtain the SD effect in device D at 83 K, corresponding to $99\%T_c$, whereas spontaneous TRSB is expected to occur at temperatures well below $50\%T_c$ at this twist angle of 40°. Thirdly, the SD effect occurs in samples with twist angles of 25° and 30° at high temperatures (as high as 77 K, device B). The diode efficiency also stays high. In these samples, we also observe half-integer Shapiro steps, reflecting the presence of second harmonic term in the CPR, at high temperatures (77 K). By contrast, the theory predicts the spontaneous TRSB at these small twist angels only at low temperatures [46] ($T < 0.1T_c$). These discrepancies indicate that the SD effect observed at relatively high temperatures in our devices has a different origin than that with spontaneous TRSB.

We employ current pulses to induce the SD effect in devices that initially host no asymmetry in the $I$-$V$ characteristics. No magnetic field is applied. This is different from the works [19,21] that employed a small perpendicular magnetic field to induce TRSB. Still, cuprates host pinning centers that can trap fluxes. The trapped fluxes, which accounts for TRSB, may likely be induced by the current pulse. They can give rise to higher harmonics in the CPR [11]. In fact, our modelling, which well captures the experimental observation (Fig. 2h), assumes the presence of a second harmonic term and an explicit TRSB term in the CPR [46]. The flux trapping is possible because Bi2212 hosts a rather small lower critical field ($B_{c1}$ ~100 µT at 50 K [47]), which could be exceeded by the self-field of a large current (Methods). The region with a twist angle

(close to 45 degrees) is most susceptible to trapped fluxes due to its relatively smaller critical Josephson current density thus smaller $B_{c1}$ [48] than that of the two constituent cuprate flakes. The absolute critical current is also much smaller across the twisted junction because of its much smaller area [19,21]. In fact, the Josephson current in devices at small twist angles, such as 15 degrees, is often too high such that current training, with an amplitude not damaging the sample, becomes less effective in inducing the SD effect. Twisting facilitates the trapping of fluxes at the particular tunnel junction under study. The involvement of trapped fluxes in our devices is further supported by the $I$-$V$ characteristics collected in device C at 70 K. There, the asymmetry can flip its sign after sweeping the current, indicating a lower energy barrier at elevated temperatures for the rearrangement of trapped fluxes.

Finally, we point out another implication of our modelling. To simulate the asymmetric $I$-$V$ characteristics (Fig. 2h), the first harmonic in the CPR is indispensable. The existence of the first order Josephson tunneling at $\theta = 45°$ challenges the theoretical understanding, because such a term should be fully suppressed in pure *d*-wave superconductors [45,46]. Its presence seems to require the assumption of an emergent *s*-wave component in the twisted cuprates [20]. Of late, the admixture of *s*-wave pairing was theoretically proposed to occur [49], as a consequence of inhomogeneity in cuprates.

In summary, we demonstrate that the QSD made of twisted cuprates meets a slew of technological demands toward practical superconducting devices: high working temperature (as high as 83 K), perfect diode efficiency ($\eta \sim 100\%$), free of any static magnetic field ($B = 0$ T), facile microwave tunability and noise filtering ability. These features promise applications in next-generation superconducting electronics.

**Methods**

**Sample fabrication**

We employed high quality $Bi_2Sr_2CaCu_2O_{8+x}$(Bi2212)/$Bi_{2-x}Pb_xSr_2CaCu_2O_{8+\delta}$ (Pb-Bi2212) single crystals with optimal doping grown by the traveling floating zone method [50,51]. The superconducting transition temperature was about 90 K/84 K. The twisted cuprate junctions were fabricated by the on-site cold stacking method developed in our previous reports [18,19]. The transfer stage mainly consisted of the sample stage and the cantilever. Each was connected to tubes for a continuous flow of cold nitrogen gas/liquid together with a heater and a thermometer. This design allowed us to control the temperatures of the cantilever and the sample stage individually. *In-situ* sample cleaving was carried out at a cryogenic temperature of -60 °C ($\pm 2$ °C). The sample release was done with the sample stage at -40 °C and the cantilever naturally warmed (nitrogen flow stopped) across a typical temperature of -20 °C. The angular precision of our devices fabricated this way is within 0.1 degree, as demonstrated from our previous transmission electron microscopy (TEM) study [19].

**Transport measurements**

After completing the stacking process inside the argon-filled glovebox, we immediately wired the device to a chip carrier within half an hour at ambient conditions and loaded the chip into a cryogenic system for measurements. The cryogenic system was equipped with a superconducting solenoid coil. To avoid trapped fluxes induced by applying current to the coil, we thermal-cycled the superconducting magnet to a temperature way above its transition temperature. Throughout the rest of the measurements, we made sure that no current was sourced to the superconducting magnet, i.e., $B = 0$ T.

The standard lock-in technique was employed to measure the junction resistance. The typical excitation current was 1 µA (13.3 Hz). The $I$-$V$ characteristics were measured by using a DC current source and a DC nanovoltmeter. We determine the instrumental noise of the DC nanovoltmeter by connecting its two inputs (A and B) and recording the voltage fluctuation over time. The FWHM of this voltage fluctuation is determined to be 0.23 μV.

To study the AC Josephson effect, a semi-rigid coaxial cable was used to channel the microwave signal from the room temperature port to the sample at cryogenic temperatures. Near the sample, the coaxial cable has an open end with the inner wire protruded out, serving as an antenna for microwave emission. We used a signal generator with a maximum frequency of 20 GHz. An amplifier (2GHz-20GHz: 20 W/43dBm) was employed to enhance the output power.

**Self-field induced by a large current**

To estimate the self-field generated by the current, we consider the magnetic field generated by a current flowing through a thin-plate: $B_y \sim \mu_0 j d$, where $\mu_0$ is the magnetic vacuum permeability, $j$ is the current density along *x*-axis, and $d$ is the thickness of the plate. We estimate $j$ based on $I_p = jwd$, where $w$ is the width of the plate and $I_p$ is the current amplitude (1.5 mA). For our estimation, we take the typical width of the flake for $w$, such that $w \sim 10$ μm. We therefore obtain: $B_y \sim \mu_0 I_p/w \sim 200$ μT. This estimated self-field is higher than the lower critical magnetic field of Bi2212 at an elevated temperature of 50 K.

**Theoretical modelling**

We describe the SD effect under microwave irradiation by adopting the scheme of ref. [14]. There, a resistively shunted junction (RSJ) model was considered with the CPR containing higher harmonic terms with TRSB. This model is suitable to describe the experimental situations obtained after applying sufficiently high current pulses (For examples, Fig. 2e-g), because the $I$ - $V$ characteristics indeed show strongly suppressed hysteresis. Specifically, we employ the CPR proposed in ref. [46]:

$$J(\varphi) = J_{c1}sin\varphi - J_{c2}sin2\varphi + J_m cos\varphi, \qquad (1)$$

where $J_{c1}$ and $J_{c2}$ are the first and second harmonics, $J_m cos\varphi$ is the term representing explicit TRSB[46]. The temporal evolution of the phase difference is[14]:

$$\frac{d\varphi}{dt} = \frac{2eR_n}{\hbar}\left(J_{dc} + J_{ac}\sin(2\pi f_{mw}t) - J(\varphi)\right), \tag{2}$$

where $R_n$ is the normal state resistance, $J_{ac}$ and $f_{mw}$ are the amplitude and frequency of an AC driving current. We numerically solve equations (1) and (2) to obtain the time dependent phase difference at a given set of $J_{dc}$, $J_{ac}$. The DC voltage output is then obtained by taking the time average following: $V_{dc} = \frac{\hbar}{2e}\langle\frac{d\varphi}{dt}\rangle$, where the arrow brackets indicate the time average operation. To simulate the experimental results obtained with different current pulses ($I_p$), we systematically vary $J_{c2}/J_{c1}$ in the CPR while maintaining $J_m = J_{c1}$. The parameters employed for generating the plots of Fig. 2h.

## Acknowledgements

This work is financially supported by the Ministry of Science and Technology of China No. 2022YFA1403100 (DZ, QKX); the National Natural Science Foundation of China [Grants No. 12141402 (YZ), 52388201 (DZ, QKX), 12361141820 (DZ), 12274249 (DZ), T2425009 (DZ)]; Innovation Program for Quantum Science and Technology [Grants No. 2021ZD0302600 (YZ, ZL), 2021ZD0302400 (DZ)]; the China Postdoctoral Science Foundation [Grant No. GZB20240294 (HW), 2024M751287 (HW). The work at BNL was supported by the US Department of Energy, office of Basic Energy Sciences, Contract No. DE-SC0012704.

## Author contributions

H. W. and Y. Z. fabricated the devices. H. W., Y. Z., and D. Z. carried out the transport measurements with technical assistance of Z. L. G. G., J. Y., L. Z., and X. J. Z. grew the single crystals. H. W. carried out the theoretical modelling. H. W., Y. Z., D.Z. and Q.-K. X. analyzed the data and wrote the paper with the input from Z. B. All authors discussed the results and commented on the manuscript.